\documentclass{cimento}

\usepackage{graphicx}
\title{LFU ratios in B decays using Lattice QCD and Unitarity}
\author{G.~Martinelli\from{ins:x}\ETC,
M.~Naviglio\from{ins:y},
S.~Simula\from{ins:w}
        \atque
L.~Vittorio\from{ins:z}\from{ins:zz}\thanks{Speaker}}
\instlist{\inst{ins:x} Physics Department, University of Roma ``La Sapienza'' and INFN, Sezione di Roma,\\ Piazzale Aldo Moro 5, 00185 Roma, Italy
  \inst{ins:y} Dipartimento di Fisica dell'Universit\`a di Pisa and INFN, Sezione di Pisa,\\ Largo Bruno Pontecorvo 3, I-56127 Pisa, Italy
  \inst{ins:w} Istituto Nazionale di Fisica Nucleare, Sezione di Roma Tre,\\ Via della Vasca Navale 84, I-00146 Rome, Italy
  \inst{ins:z} Scuola Normale Superiore,\\ Piazza dei Cavalieri 7, 56126 Pisa, Italy
   \inst{ins:zz} Istituto Nazionale di Fisica Nucleare, Sezione di Pisa,\\ Largo Bruno Pontecorvo 3, I-56127 Pisa, Italy}
\begin{document}

\maketitle

\begin{abstract}
We present the results of the application of the unitarity-based Dispersion Matrix approach to semileptonic charged-current $B$ decays. This method allows to achieve a non-perturbative and completely model-independent determination of the hadronic form factors. Starting from lattice results available at large values of the momentum transfer, the behaviour of the form factors in their whole kinematical range is obtained without introducing any explicit parameterization of their momentum dependence. We will focus on the analysis of Lepton Flavour Universality by computing the $\tau/\mu$ ratios of the branching fractions of the $B \to D^{(*)} \ell \nu$ and $B \to \pi \ell \nu$ decays. The most important result is that, for the first time, the discrepancies between the SM expectation values and the measurements of the Lepton Flavour Universality ratios for the $B \to D^{(*)} \ell \nu$ decays are reduced at the 1.3$\sigma$ level for each of the two channels, separately.
\end{abstract}

\section{State-of-the-art of exclusive $B \to D^{(*)} \ell \nu$ decays}

$B \to D^{(*)} \ell \nu$ decays are among the most challenging processes in the phenomenology of flavor physics, since they are affected by two unsolved problems. 

On the one hand, we have the so-called $\vert V_{cb} \vert$ \emph{puzzle}, $i.e.$ the discrepancy between the inclusive and the exclusive determinations of the CKM matrix element $\vert V_{cb}\vert$. According to the FLAG Review 2021 \cite{FLAG21}, there is a $\sim2.8\sigma$ tension between the exclusive estimate (that depends on the form factors parametrization) and the inclusive one, namely
\begin{equation}
\label{VcbFLAG21}
\vert V_{cb} \vert_{\rm excl} \times 10^3 = 39.36(68),\,\,\,\,\,\,\,\,\,\,\,\,\vert V_{cb} \vert_{\rm incl} \times 10^3 = 42.00(65).
\end{equation}
A new more precise estimate of the inclusive value has also recently appeared \cite{Bordone:2021oof}, namely $\vert V_{cb} \vert_{\rm incl}=42.16(50)$, which is compatible with the inclusive FLAG value in Eq.\,(\ref{VcbFLAG21}). 

On the other hand, a strong tension exists between the theoretical value and the measurements of $R(D^{(*)})$, which are a fundamental test of Lepton Flavour Universality (LFU) and are defined as 
\begin{equation}
\label{RDdef}
R(D^{(*)}) \equiv \frac{\Gamma(B \to D^{(*)} \tau \nu_{\tau})}{\Gamma(B \to D^{(*)} \ell \nu_{\ell})},
\end{equation}
where $\ell$ denotes a light lepton. The HFLAV Collaboration \cite{HFLAV} computed the world averages of the available measurements of the $R(D^{(*)})$ ratio and of their SM theoretical predictions. From the numerical point of view, we have 
\begin{equation}
\label{RDHFLAV}
R(D)_{\rm SM} = 0.299 \pm 0.003,\,\,\,\,\,R(D)_{\rm exp} = 0.339 \pm 0.026 \pm 0.014
\end{equation}
for the $B \to D$ case and 
\begin{equation}
\label{RDstHFLAV}
R(D^*)_{\rm SM} = 0.254 \pm 0.005,\,\,\,\,\,R(D)_{\rm exp} = 0.295 \pm 0.010 \pm 0.010
\end{equation}
for the $B \to D^*$ one. As clearly stated by HFLAV Collaboration, the averages of the measurements of $R(D)$ and $R(D^*)$ exceed the corresponding SM predictions by 1.4$\sigma$ and 2.8$\sigma$, respectively. If we also take into consideration the experimental correlation between these two quantities, namely $\rho=-0.38$, the resulting difference with the SM predictions is increased at the 3.3$\sigma$ level.

%\begin{figure}
%\label{RDHFLAV}
%\includegraphics{rdrds_2021}
%\caption{\textit{Correlation plot of $R(D)$-$R(D^*)$ as reported by the HFLAV Collaboration \cite{} in the Spring 2021. The red area represents the world average of all the available measurements, while the black cross is the average of the SM expectation values. The plot makes evident a 3.3$\sigma$ tension between theory and experiments.}
%\hspace*{\fill} \small}
%\end{figure}

\section{The Dispersion Matrix approach to Form Factors in $B \to D^{(*)} \ell \nu$ decays}

Let us focus on semileptonic $B \to D^{(*)}$ decays. In case of production of a \emph{pseudoscalar} meson, $i.e.$ the $B \to D \ell \nu$ case, the differential decay width reads
\begin{equation}
\label{finaldiff333}
\frac{d\Gamma}{dq^2}=\frac{G_F^2 \vert V_{cb} \vert^2 \eta_{EW}^2}{24\pi^3} \left(1-\frac{m_{\ell}^2}{q^2}\right)^2 \times \left[ C_+(q^2)+C_0(q^2) \right],
\end{equation}
where
\begin{eqnarray}
C_+(q^2)&\equiv& \vert \vec{p}_{D}\vert^3 \left(1+\frac{m_{\ell}^2}{2q^2}\right) \vert f^+(q^2) \vert^2,\\
C_0(q^2)&\equiv& m_{B}^2 \vert \vec{p}_{D} \vert \left( 1-\frac{m_{D}^2}{m_{B}^2}\right)^2 \frac{3m_{\ell}^2}{8q^2} \vert f^0(q^2) \vert^2.
\end{eqnarray}
In this case only two Form Factors (FFs) are present, namely $f^+(q^2),\,f^0(q^2)$. Moreover, $\vec{p}_{D}$ represents the 3-momentum of the produced $D$ meson and $m_{\ell}$ is the mass of the produced lepton.

In case of production of a \emph{vector} meson, $i.e.$ the $B \to D^* \ell \nu$ case, the expression of the differential decay width is 
\begin{equation}
\label{finaldiff333D*}
\frac{d\Gamma_{\tau}}{dw} = \frac{d\Gamma_{\tau,1}}{dw} + \frac{d\Gamma_{\tau,2}}{dw},
\end{equation}
where
\begin{eqnarray}
\frac{d\Gamma_{\tau,1}}{dw} &=&\frac{\eta_{EW}^2  \vert V_{cb} \vert^2 G_F^2 m_{D^*}^2}{48 \pi ^3 m_{B}} \sqrt{w^2-1} \left( 1-\frac{m_{\tau}^2}{q(w)^2} \right)^2 \left( 1+\frac{m_{\tau}^2}{2 q(w)^2} \right)  \,K(w),\\
K(w)&\equiv& 2 \,q^2(w) \left(f(w)^2+
   m_{B}^2 m_{D^*}^2 \left(w^2-1\right)g(w)^2\right)+\mathcal{F}_1(w)^2,\\
   \frac{d\Gamma_{\tau,2}}{dw} &=& \frac{\eta_{EW}^2 \vert V_{cb} \vert^2 G_F^2 m_B^5}{32 \pi^3}\frac{m_{\tau}^2(m_{\tau}^2-q(w)^2)^2 r^3(1+r)^2(w^2-1)^{3/2} P_1(w)^2}{q(w)^6}.
\end{eqnarray}
In this case we have four FFs to deal with, $i.e.$ $f(w),g(w),\mathcal{F}_1(w),P_1(w)$. Note that in Eqs.\,(\ref{finaldiff333}) and (\ref{finaldiff333D*}) we refer equivalently to the momentum transfer $q^2$ or to the recoil $w$, since they are related by the following 1-to-1 correspondence
\begin{equation}
q^2(w)= m_B^2+m_{D^{(*)}}^2 - 2 m_B m_{D^{(*)}} w.
\end{equation}

Now, our goal is to describe the FFs entering in $B \to D^{(*)} \ell \nu$ decays by using the novel Dispersion Matrix (DM) method \cite{DiCarlo:2021dzg}, which was originally proposed in \cite{Lellouch96}. The DM method allows us to study the FFs in a non-perturbative and model-independent way, since, starting from the available LQCD computations of the FFs at high momentum transfer (or, equivalently, at low recoil), we can extrapolate their behaviour in the opposite kinematical region. To this end, we do not assume any functional dependence of the FFs on $q^2$ (or, equivalently, on $w$) and we use only non-perturbative inputs. Moreover, the resulting bands of the FFs will be independent of the experimental determinations of the differential decay widths. 

From the mathematical point of view, the starting point is to focus on one of the six FFs defined above, for instance $f$, and then consider the matrix
\begin{equation}
\label{eq:Delta2}
%\centering
\mathbf{M} = \left( 
\begin{tabular}{ccccc}
   $\chi$ & $\phi f$ & $\phi_1 f_1$ & $...$ & $\phi_N f_N$ \\[2mm] 
   $\phi f$ & $\frac{1}{1 - z^2}$ & $\frac{1}{1 - z z_1}$ & $...$ & $\frac{1}{1 - z z_N}$ \\[2mm]
   $\phi_1 f_1$ & $\frac{1}{1 - z_1 z}$  & $\frac{1}{1 - z_1^2}$ & $...$ & $\frac{1}{1 - z_1 z_N}$ \\[2mm]
   $... $  & $...$ & $...$ & $...$ & $...$ \\[2mm]
   $\phi_N f_N$ & $\frac{1}{1 - z_N z}$ & $\frac{1}{1 - z_N z_1}$ & $...$ & $\frac{1}{1 - z_N^2}$
\end{tabular}
\right),
\end{equation}
where we have introduced the conformal variable $z$ defined as 
\begin{equation}
\label{conf}
z(t) = \frac{\sqrt{t_+ -t} - \sqrt{t_+ - t_-}}{\sqrt{t_+ -t} + \sqrt{t_+ - t_-}},\,\,\,\,\,\,\,\,\,\,\,\,\,\,\,\,\,\,\,\,\,t_{\pm}=(m_B \pm m_{D^{(*)}})^2
\end{equation}
or, equivalently, as
\begin{equation}
z=\frac{\sqrt{w+1}-\sqrt{2}}{\sqrt{w+1}+\sqrt{2}}.
\end{equation}
In this expression, $\phi_i f_i \equiv \phi(z_i) f(z_i)$ (with $i = 1, 2, ... N$) represent the known values of the quantity $\phi(z) f(z)$ corresponding to the values $z_i$ at which the FFs have been computed on the lattice. The kinematical function $\phi(z)$ has a specific expression for each of the aforementioned FFs. The general forms for each case can be found in \cite{paperoIII}. Finally, the susceptibility $\chi(q^2)$ is related to the derivative with respect to $q^2$ of  the Fourier transform of suitable Green functions of bilinear quark operators and follows from the dispersion relation associated to a particular spin-parity quantum channel. Note that they have been computed for the first time on the lattice in \cite{paperoII} for $b \to c$ quark transitions. At this point, one can demonstrate from first principles that $\det \mathbf{M} \geq 0$. Then, the positivity of the determinant, which we will refer to as \emph{unitarity filter} hereafter, allows to compute the lower and the upper bounds of the FF of interest for each generic value of $z$, $i.e.$
\begin{equation}
f_{\rm lo}(z) \leq f(z) \leq f_{\rm up}(z).
\end{equation}
The explicit definitions of $f_{\rm lo}(z),\,f_{\rm up}(z)$ can be found in \cite{DiCarlo:2021dzg}.

\section{An instructive example: the semileptonic $B \to D^*$ channel}

Let us discuss in detail the semileptonic $B \to D^*$ decay, which is very challenging due to the high number of FFs involved. In \cite{EPJC} we have computed the unitarity bands of the FFs, starting from the final results of the computations on the lattice performed by the FNAL/MILC Collaborations \cite{FNALMILCD*}. There, in the ancillary files, the authors give the synthetic values of the FFs $g(w), f(w), \mathcal{F}_1(w)$ and $\mathcal{F}_2(w)$ at three non-zero values of the recoil variable ($w-1$), namely $w = \{1.03,1.10,1.17\}$, together with their correlations. Note that the FF $\mathcal{F}_2(w)$ is directly related to the $P_1(w)$ one, in fact $P_1(w) =  \mathcal{F}_2(w) \sqrt{r} / (1 + r)$, where $r \equiv m_{D^*}/m_B \simeq 0.38$.

%\begin{figure}[h!]
%\begin{center}
%\includegraphics[scale=0.5]{fig1.pdf}
%\caption{\it \small The bands of the FFs $g(w)$, $f(w)$, $\mathcal{F}_1(w)$ and $P_1(w)$ computed by the DM method after imposing both the unitarity filter and the two KCs~(\ref{eq:KC1})-(\ref{eq:KC2}). The FNAL/MILC values~\cite{FermilabLattice:2021cdg} used as inputs for the DM method are represented by the black diamonds.}
%\label{FFD*}
%\end{center}
%\end{figure}

\begin{figure}
\centering
\includegraphics[width=.8\textwidth]{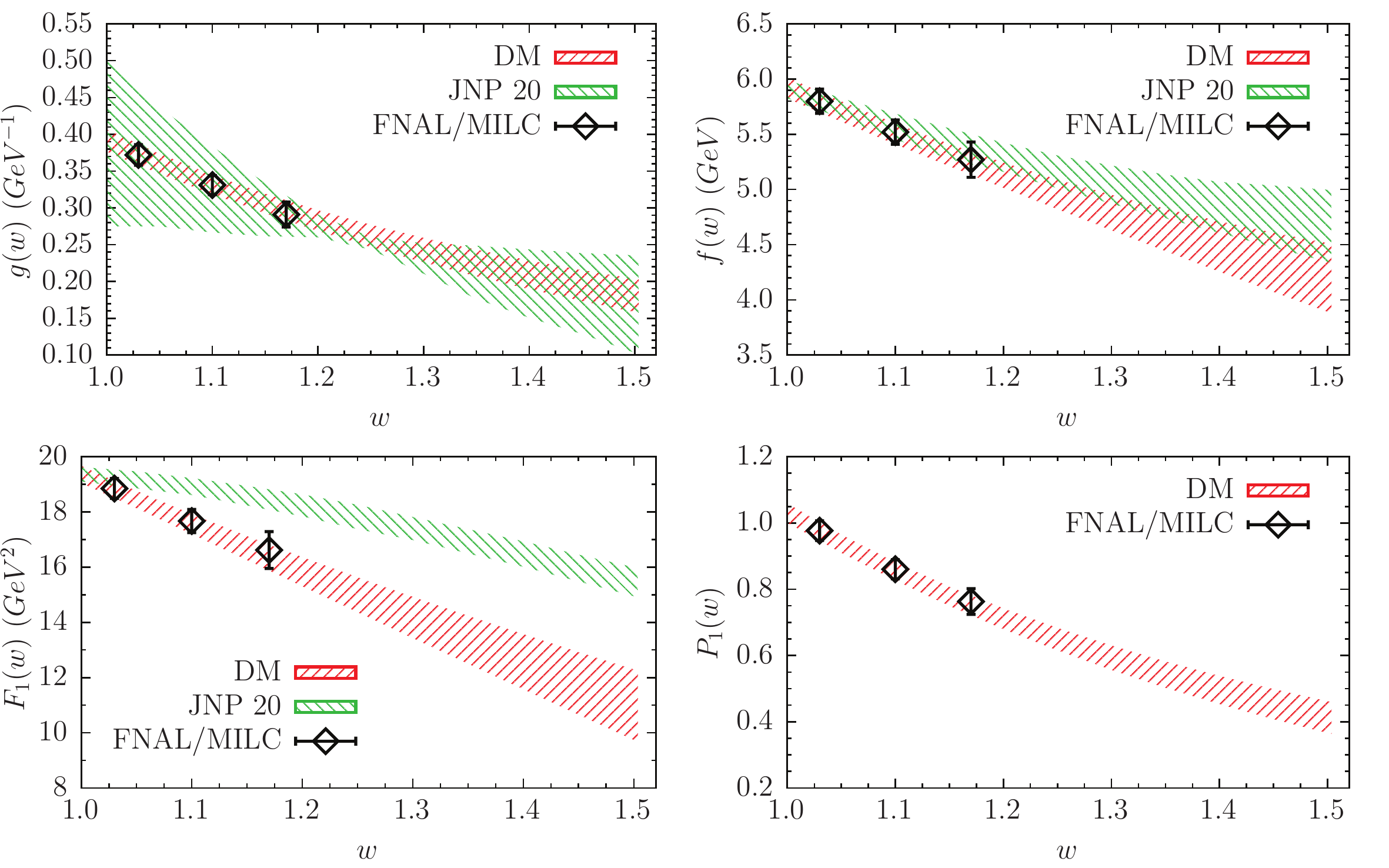}
\caption{\small The red bands of the FFs $g(w)$, $f(w)$, $\mathcal{F}_1(w)$ and $P_1(w)$ are computed through the DM method after imposing both the unitarity filter and the two KCs~(\ref{eq:KC1})-(\ref{eq:KC2}). The FNAL/MILC values~\cite{FNALMILCD*} used as inputs for the DM method are represented by the black diamonds. The green bands are, instead, the FFs obtained by fitting the Belle experimental data \cite{Venezia} for the differential decay widths. These bands have been taken from Fig.\,2 of \cite{JNP}.}
\label{FFD*bis}
\end{figure}

In Fig.\,\ref{FFD*bis} we show the results of our DM analysis as red bands. The DM unitarity bands are built up through bootstrap events that satisfy exactly both the unitarity filter of the DM method and the Kinematical Constraints (KCs) 
\begin{eqnarray}
    \label{eq:KC1}
    \mathcal{F}_1(1) & = & m_B (1 - r) f(1),\\
    \label{eq:KC2}
     P_1(w_{max}) & = & \frac{\mathcal{F}_1(w_{max})}{m_B^2 (1 + w_{max}) (1 -r) \sqrt{r}},
\end{eqnarray}
where $w_{\rm max} \simeq 1.5$. These unitarity bands allow us to compute new \emph{fully-theoretical} values of the LFU ratio $R(D^*)$, thus obtaining
\begin{equation}
\label{RDstDM}
R(D^*) = 0.275 \pm 0.008.
\end{equation}
For the first time, the compatibility between this theoretical determination of $R(D^*)$ and the corresponding HFLAV world average of the measurements is at the $\sim1.3 \sigma$. 

At this point, a natural question arises: why is the SM HFLAV theoretical average in Eq.\,(\ref{RDstHFLAV}) so different from the DM estimate in Eq.\,(\ref{RDstDM})? The answer has to be found in the shape of the FFs. In Fig.\,\ref{FFD*bis} the red DM unitarity bands are compared with new green ones, taken directly from Fig.\,2 of \cite{JNP}. In this article, the authors show the plot of the FFs $g(w)$, $f(w)$ and $\mathcal{F}_1(w)$ obtained by fitting the Belle experimental data \cite{Venezia} for the differential decay widths through the BGL parametrization \cite{BGL1, BGL2, BGL3}. This comparison is particularly instructive since from the green bands the authors of \cite{JNP} obtain the result $R(D^*) = 0.251^{+0.004}_{-0.005}$, which is very similar to the SM HFLAV result (\ref{RDstHFLAV}). To achieve this goal, the authors of \cite{JNP} had also constrained the pseudoscalar FF $P_1(w)$ through appropriate Heavy Quark Effective Theory relations, as described in detail in \cite{JNPold}. Figure \ref{FFD*bis} clearly shows that the different shapes of the FF $\mathcal{F}_1(w)$ induced by the LQCD or the experimental data have a fundamental impact on the final theoretical value of $R(D^*)$. To avoid the bias induce by the experimental data on $R(D^*)$, we have then adopted the DM method by using \emph{only} the LQCD computations as inputs for the description of the unitarity bands of the FFs.

\section{A summary of the DM applications to all the $b \to c$ quark transitions}

\begin{figure}
\centering
\includegraphics[width=.7\textwidth]{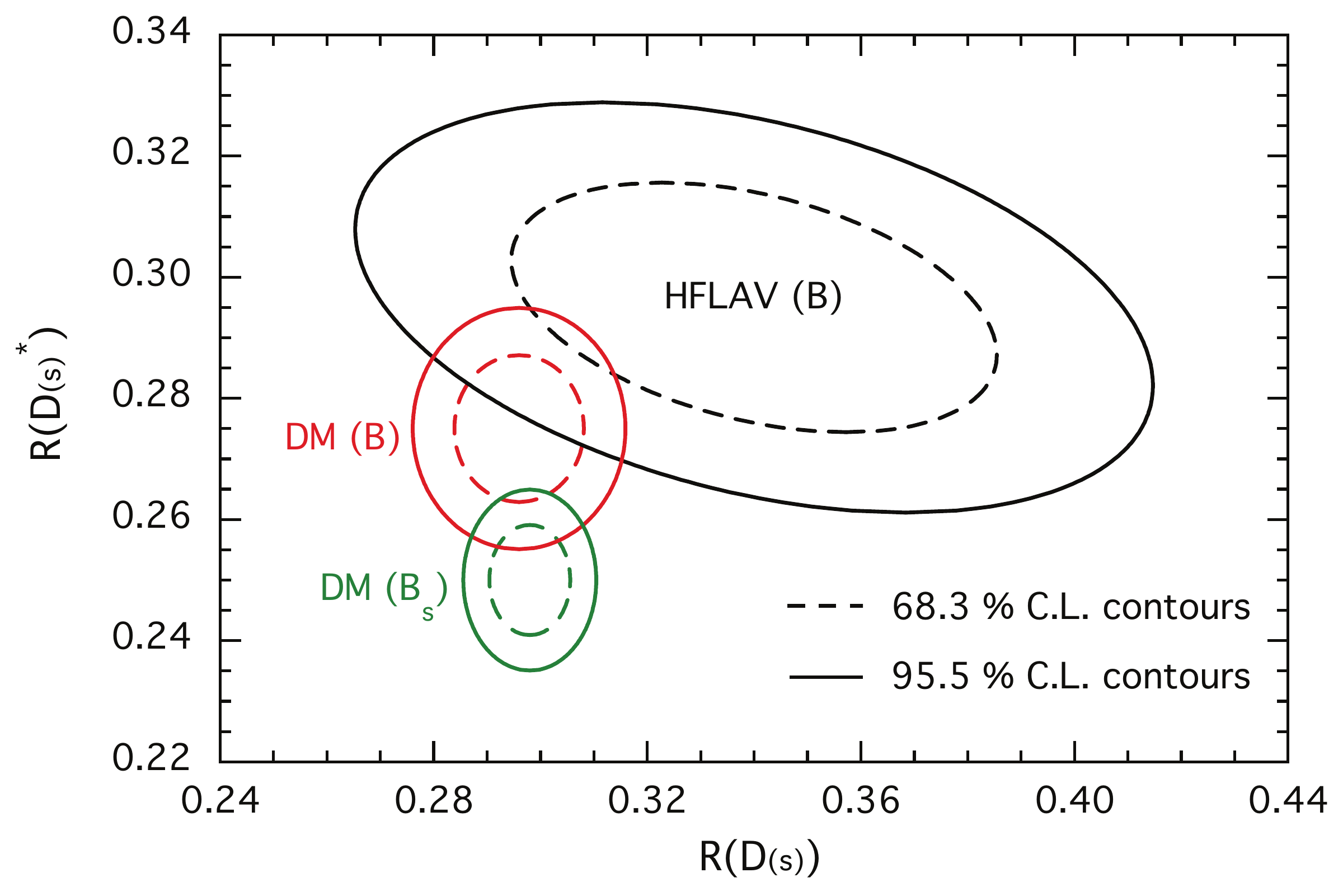}
\caption{\small The correlation plot for $R(D_{(s)})$ and $R(D_{(s)}^*)$. The black area represents the average of all the experimental measurements of $R(D)$ and $R(D^*)$ performed by HFLAV Collaboration. The red and the green regions are the DM predictions for $R(D_{(s)})$ and $R(D_{(s)}^*)$.}
\label{ellissiLFU}
\end{figure}

The application of the Dispersive Matrix method, that we have explicitly described in the case of semileptonic $B \to D^*$ decays, can be repeated to all the other relevant $b \to c$ quark transitions in the mesonic sector, namely the $B \to D \ell \nu$, the $B_s \to D_s \ell \nu$ and the $B_s \to D_s^* \ell \nu$ processes. In the former case, in \cite{paperoIII} we have obtained the unitarity bands of the FFs $f_+,\,f_0$, defined in Eq.\,(\ref{finaldiff333}), by using the FNAL/MILC synthetic data in \cite{FNALMILCD}. In the latter two cases, in \cite{PaperDs} we have extracted three values of the relevant FFs from the fits published by the HPQCD Collaboration in \cite{HPQCD1, HPQCD2} by using the BCL parametrization \cite{BCL} and we have then implemented the DM method. In Fig.\,\ref{ellissiLFU} we show the DM estimates of all the LFU observables characterizing these decays, $i.e.$ the $R(D_{(s)})$ and $R(D_{(s)}^*)$ ratios. The black area represents the average of all the experimental measurements of $R(D)$ and $R(D^*)$, computed by HFLAV and reported in Eqs.\,(\ref{RDHFLAV})-(\ref{RDstHFLAV}). The red and the green regions are, instead, the DM predictions for $R(D_{(s)})$ and $R(D_{(s)}^*)$ obtained in Refs.\,\cite{paperoIII, EPJC, PaperDs}. 

Our first message is that the anomalies in the $B$ sector have been very lightened through the DM approach to the hadronic FFs. This achievement is mainly due to the absence of the mixing between theoretical and experimental data in the description of the FFs. Our second message is that the comparison between $R(D^*)$ and $R(D_s^*)$, which differ by $\approx 10\%$, highlights the possible presence of $SU(3)_F$ symmetry breaking effects in semileptonic charged-current $B$ decays. In this sense, further studies of the spectator-quark dependence of the hadronic FFs of the semileptonic $B \to D^{(*)}$ and  $B_s \to D_s^{(*)}$ transitions are called for. 

\section{Lepton Flavour Universality in semileptonic $B \to \pi$ decays}

The DM method can be applied to whatever semileptonic charged-current decays of mesons and baryons. It is very instructive, thus, to investigate its potential in the analysis of the $b \to u$ quark transitions.

Let us discuss, for instance, the case of the $B \to \pi \ell \nu$ decays \cite{paperoV}. They are characterized by two FFs, which are analogous to the ones defined in Eq.\,(\ref{finaldiff333}), given the pseudoscalar nature of the $\pi$ meson. Let us call them $f_+^{\pi},\,f_0^{\pi}$ to distinguish them from the $B \to D$ case. These FFs have been studied by the RBC/UKQCD\,\cite{Flynn:2015mha} and the FNAL/MILC\,\cite{Lattice:2015tia} Collaborations. For both channels the lattice computations of the FFs are available in the large-$q^2$ region. To be more specific, the authors of Ref.\,\cite{Flynn:2015mha} provide synthetic LQCD values of the FFs (together with their statistical and systematic correlations) at $q^2 = \{ 19.0, 22.6, 25.1 \}$ GeV$^2$. In \cite{Lattice:2015tia}, instead, only the results of BCL fits of the FFs extrapolated to the continuum limit and to the physical pion point are available. Thus, from the marginalized BCL coefficients we evaluate the mean values, uncertainties and correlations of the FFs at the same three values of $q^2$ given in Ref.\,\cite{Flynn:2015mha}. 

\begin{figure}
\centering
\includegraphics[width=.8\textwidth]{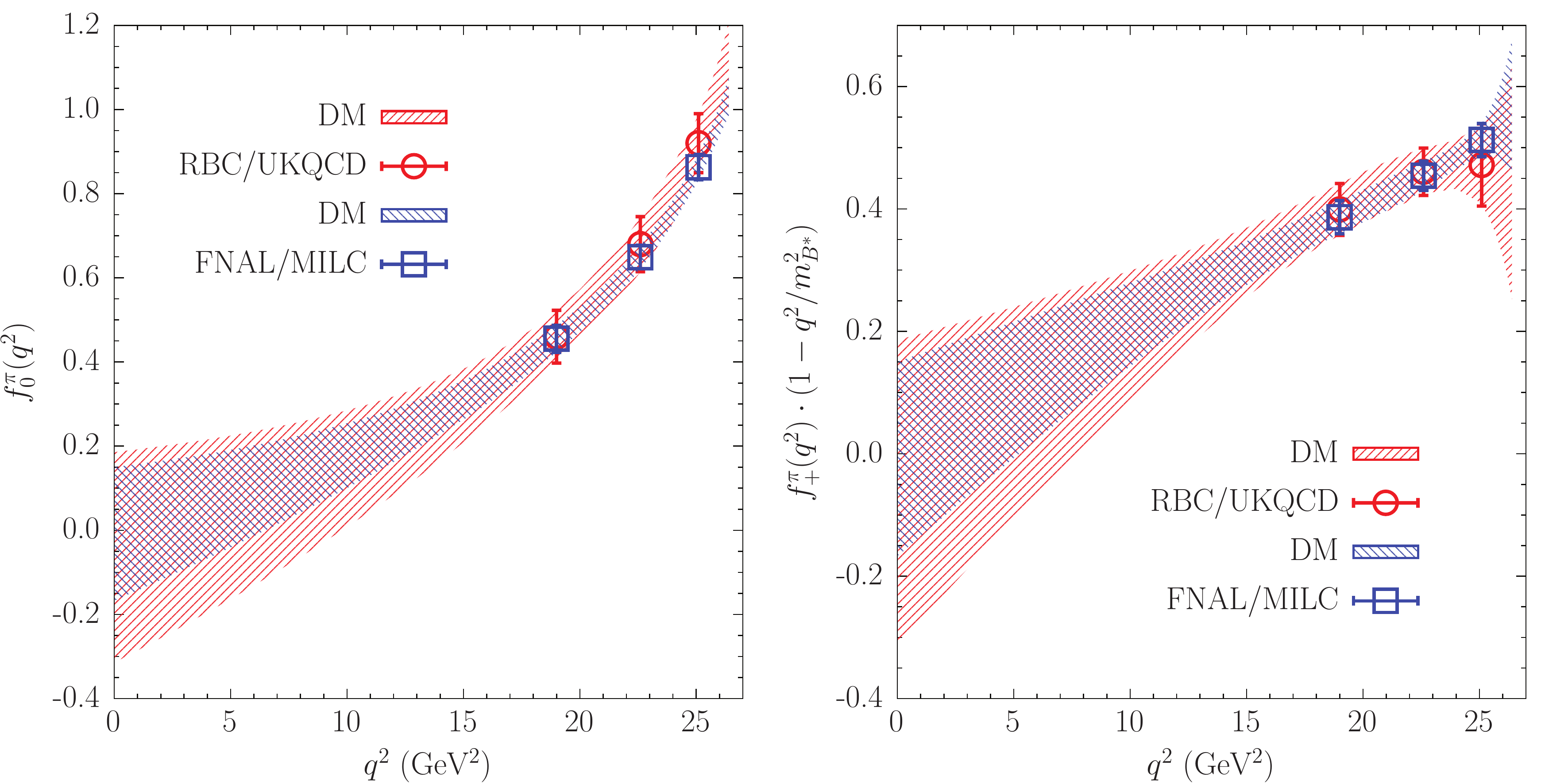}
\caption{\small The scalar $f_0^\pi(q^2)$ (left panel) and vector $f_+^\pi(q^2)$ (right panel) FFs entering the semileptonic $B \to \pi \ell \nu_\ell$ decays computed by the DM method as a function of the 4-momentum transfer $q^2$ using the LQCD inputs from RBC/UKQCD (red points) and FNAL/MILC (blue squares) Collaborations. In the right panel, the vector FF is multiplied by the factor $(1 - q^2 / m_{B^*}^2)$ with $m_{B^*} = 5.325$ GeV.\hspace*{\fill}}
\label{FFsBpi}
\end{figure}

In Fig.\,\ref{FFsBpi} we show the red (blue) DM bands which are obtained by using as inputs the RBC/UKQCD (FNAL/MILC) data, respectively. In principle, when one implements the BCL fits the mean value and the uncertainty of the FFs value extrapolated at zero momentum transfer are not stable under variation of the truncation order of a series expansion of the FFs. On the contrary, the DM approach is completely independent of this issue, since no approximation due to the truncation of a series expansion is present. In other words, we argue that the DM method is equivalent to the results of all possible (BCL) fits which satisfy unitarity and, at the same time, reproduce exactly the input data. Note that this property is particularly useful in $B \to \pi \ell \nu$ decays, since here we have a long extrapolation in $q^2$.

These unitarity bands can be used to compute fully-theoretical values of the LFU ratio $R_{\pi}^{\tau/\mu} \equiv \Gamma(B \to \pi \tau \nu_{\tau})/\Gamma(B \to \pi \mu \nu_{\mu})$, which is equivalent to the $R(D^{(*)})$ ratios defined in Eq.\,(\ref{RDdef}). The DM results are shown in Table \ref{tab:phenoBpi} for each LQCD input. The \emph{combined} case in the last column corresponds to the combination of the RBC/UKQCD and the FNAL/MILC data, as described in \cite{paperoV}. They are all compatible with the only available measurement by Belle \cite{Hamer:2015jsa}\begin{equation}
\label{RBelle}
R_{\pi}^{\tau/\mu}\vert_{exp} = 1.05 \pm 0.51,
\end{equation}
which has a large uncertainty compared to theoretical predictions. However, the expected uncertainty on the above ratio by Belle II at 50 ab$^{-1}$ of luminosity\,\cite{BelleII} is $\delta R_{\pi}^{\tau/\mu} \simeq 0.09$, comparable to our present theoretical uncertainties. Further LQCD computations of the FFs, as well as more precise ones, will thus be of capital importance in order to test possible New Physics effects affecting semileptonic $B \to \pi$ decays.

\begin{table}[h!]
%\begin{center}
\begin{tabular}{rccc}
\hline
Input & RBC/UKQCD & FNAL/MILC & combined\\
\hline
$R^{\tau/\mu}_{\pi}$ & 0.767(145) & 0.838(75) & 0.793(118) \\
\hline
\end{tabular}
\caption{\it Theoretical values of $R_{\pi}^{\tau/\mu}$ in the case of semileptonic $B \to \pi$ decays adopting the RBC/UKQCD, the FNAL/MILC and the combined LQCD data as inputs for our DM method.\hspace*{\fill}}
\label{tab:phenoBpi}
%\end{center}
\end{table}

\section{Conclusions}

We have reviewed the main properties of the Dispersion Matrix approach, which is an attractive tool to implement unitarity and LQCD calculations in the analysis of exclusive semileptonic decays of mesons and baryons. It has several interesting properties. In particular, it does not rely on any assumption about the momentum dependence of the hadronic form factors and it can be based entirely on first principles ($i.e.$ unitarity and analiticity) and on non-perturbative inputs. 

We have discussed the application of the DM method to the $b \to c$ and the $b \to u$ quark transitions and the resulting theoretical determinations of the LFU observables. In this sense, our main result is that the anomalies in semileptonic charged-current $B$ decays have been strongly lightened. In fact, we have consistency between theory and experiment at the $1.3\sigma$ level for both $R(D)$ and $R(D^*)$, separately. To achieve this goal, it is fundamental to avoid the mixing among theoretical calculations and experimental data to describe the shape of the FFs of interest.

%\subsection{Mathematics}
%Here is a lettered array~(\ref{e.all}), with eqs.~(\ref{e.house})
%and~(\ref{e.phi}):
%\begin{eqnletter}
% \label{e.all}
% \drm x_\sy{F} & = & 1.2\cdot10^3\un{cm}, \qquad
%                     \tx{where\ } \sy{F} = \tx{Fermi}    \label{e.house}\\
% \phi_i        & = & i\pi                                \label{e.phi}
%\end{eqnletter}

%%\subsection{Citations}
%We're almost done, just some citations~\cite{ref:apo}
%and we will be over~\cite{ref:pul,ref:bra}.

%\acknowledgments
%L.V. warmly acknowledges Gino Isidori and all the Organizers of \emph{La Thuile 2022 - XXXV Rencontres de Physique de La Vall\'ee d'Aoste} conference for having given him the opportunity to present the results contained in this Proceeding. 

\end{document}